\documentclass[twocolumn,10pt,cleanfoot, cleanhead]{asme2e}

\usepackage{amsmath}
\usepackage{amsfonts}
\usepackage{graphicx}
\usepackage{amssymb}
\usepackage{hyperref}
\usepackage{algorithm}
\hypersetup{
    colorlinks=true,
    linkcolor=blue,
    filecolor=magenta,      
    urlcolor=cyan,
}

\confshortname{IDETC/CIE 2019}
\conffullname{the ASME 2019 International Design Engineering Technical Conferences \&\\
              Computers and Information in Engineering Conference}

\confdate{August 18-21}
\confyear{2019}
\confcity{Anaheim, CA}
\confcountry{USA}

\papernum{IDETC2019-98099}

\title{Deep active subspaces - a scalable method for high-dimensional uncertainty propagation}

\author{Rohit Tripathy
    \affiliation{
	School of Mechanical Engineering\\
	Purdue University\\
	West Lafayette, Indiana 47906\\
    Email: rtripath@purdue.edu
    }	
}

\author{Ilias Bilionis\thanks{Address all correspondence to this author.} 
    \affiliation{
    School of Mechanical Engineering\\
	Purdue University\\
	West Lafayette, Indiana 47906\\
    Email: ibilion@purdue.edu
    }
}

\newcommand{\btheta}{\boldsymbol{\theta}}
\newcommand{\bphi}{\boldsymbol{\phi}}
\newcommand{\X}{\mathbf{X}}

\newcommand{\bx}{\mathbf{x}}
\newcommand{\y}{\mathbf{y}}
\newcommand{\z}{\mathbf{z}}
\newcommand{\W}{\mathbf{W}}
\newcommand{\R}{\mathbb{R}}

\newcommand{\calD}{\mathcal{D}}

\newcommand{\calN}{\mathcal{N}}

\newcommand{\bbeta}{\boldsymbol{\beta}}
\newcommand{\bGamma}{\boldsymbol{\Gamma}}
\newcommand{\bxi}{\boldsymbol{\xi}}
\newcommand{\bzeta}{\boldsymbol{\zeta}}
\newcommand{\bXi}{\boldsymbol{\Xi}}

\begin{document}

\maketitle    

\begin{abstract}
\label{sec:abstract}
{\it 
A problem of considerable importance within the field of uncertainty quantification (UQ) is the development of efficient methods for the construction of accurate surrogate models. 
Such efforts are particularly important to applications constrained by high-dimensional uncertain parameter spaces.
The difficulty of accurate surrogate modeling in such systems, is further compounded by data scarcity brought about by the large cost of forward model evaluations. 
Traditional response surface techniques, such as Gaussian process regression (or Kriging) and polynomial chaos are difficult to scale to high dimensions.
To make surrogate modeling tractable in expensive high-dimensional systems, one must resort to dimensionality reduction of the stochastic parameter space. 
A recent dimensionality reduction technique that has shown great promise is the method of `active subspaces'. 
The classical formulation of active subspaces, unfortunately, requires gradient information from the forward model - often impossible to obtain. 
In this work, we present a simple, scalable method for recovering active subspaces in high-dimensional stochastic systems, without gradient-information that relies on a reparameterization of the orthogonal active subspace projection matrix, and couple this formulation with deep neural networks.
We demonstrate our approach on synthetic and real world datasets and show favorable predictive comparison to classical active subspaces.  
}
\end{abstract}

\begin{nomenclature}
\entry{$\mathbf{W}$}{Active subspace projection matrix}
\entry{$g$}{Link function}
\entry{$\bxi$}{Stochastic parameters}
\entry{AS}{Active subspace}
\end{nomenclature}

\section{Introduction}
\label{sec:introduction}
Inspite of the advent of the exascale era of computing and the rapid increase in the availability of computational resources \cite{reed2015exascale}, the sophistication of computer codes that simulate physical systems have also risen exponentially, either due to the incorporation of more realistic physics or higher-order numerical algorithms. 
A further consequence of the increasing sophistication of modern day computational solvers is the introduction of more parameters into the model to accurately describe boundary/initial conditions, material properties, constitutive laws etc. 
It is often the case that many (or all) of these parameters and unknown exactly. 
This brings up several questions for the computational scientist such as - i. how must one go about making robust predictions about the quantities of interest in a complex simulation, ii. how can one assess the impact of input parameter uncertainty on the model outputs, iii. how can one calibrate the model from experimental data, and so on. 
Answering such questions lie at the heart of uncertainty quantification (UQ) \cite{smith2013uncertainty, sullivan2015introduction}. 

The most common task in UQ is what is known as the forward UQ or uncertainty propagation (UP) problem \cite{le2004uncertainty, knio2006uncertainty, lee2009comparative}. A complete description of the UP problem can be found in Sec. \ref{sec:up}. 
UP is the task of estimating the statistical properties of the model outputs given a formal description of the uncertainty in the model input parameters. 
Monte Carlo (MC) methods \cite{robert2013monte} are the most straightforward techniques for solving the UP problem and have, indeed, long been the workhorse of UQ \cite{watt2012study, baraldi2008combined, rochman2014efficient}. 
A remarkable property of MC is that the variance of the standard MC estimate converges independent of the dimensionality of the stochastic parameters \cite{robert2013monte}. However, MC methods require a very large number of samples (hundreds of thousands) to show convergence in statistics \cite{robert2013monte}. 
This makes MC infeasible for state-of-the-art numerical simulators which have a large computational burden associated to each individual run. 

To overcome the fact that one does not have the computational budget for hundreds of thousands of runs of a numerical simulator, one resorts to the surrogate approach to UP \cite{nobari2015uncertainty}. 
The idea is simple - perform a limited number of forward model evaluations, collect the resulting data, and construct a cheap-to-evaluate, yet accurate, map between the input uncertain parameters and the model outputs. 
This map serves as approximation to the true solver, and is referred to as a \textit{surrogate} or a \textit{response surface}. 
Since the response surface can be repeatedly evaluated very quickly, it is now easy to couple the surrogate with a MC approach to estimate model output statistics. 
Surrogate models can be either \textit{intrusive} (i.e. requiring modification of the simulator for the analogous deterministic problem) or \textit{non-intrusive} (i.e. where the application of the surrogate model is an `outer-loop' process and the numerical solver can be treated as a black-box). 
Naturally, given the sophistication of state-of-the-art numerical models, intrusive methods such as \textit{polynomial chaos} \cite{xiu2002wiener, xiu2003modeling} have fallen out of favor in recent times. 
This has coincided with the rise in popularity of non-intrusive methods such as \textit{Gaussian process regression (GPR)} (or \textit{Kriging}) \cite{stein2012interpolation,rasmussen2003gaussian}. 

The surrogate approach to UQ has seen tremendous success in a broad range of applications \cite{zhang2012uncertainty, angelikopoulos2012bayesian, knio2006uncertainty}. 
However, state-of-the-art surrogate modeling techniques become exponentially difficult to scale as the number of stochastic parameters increases \cite{tripathy2016gaussian, constantine2014active}.
This is due to the phenomenon known, universally, as the \textit{curse of dimensionality} (CoD). 
Originally coined by Bellman in the context of dynamic programming \cite{bellman1956dynamic}, the CoD refers to the phenomenon where the volume of the input space that one must explore to gather data sufficient for constructing an accurate response surface, rises exponentially with a linear increase in the input dimensionality. 
To address the CoD, one needs to perform \textit{dimensionality reduction} of the stochastic parameter space.
The easiest approach to parameter space reduction is through a ranking of the importance of individual input dimensions. 
Methods that adopt this approach include sensitivity analysis \cite{saltelli2004sensitivity} and automatic relevance determination (ARD) \cite{neal1998assessing}. Unfortunately, such variable reduction techniques are most effective when the input variables have some degree of correlation. 
In generic UP problems, the stochastic parameters are frequently uncorrelated. 
For instance, the common scenario of functional uncertainties (such as random permeability in porous media flow) comprise of high (or infinite) dimensional stochastic parameter spaces with statistical independence between input dimensions.
The most popular approach to dimensionality reduction within the UQ community is the truncated \textit{Karhunen-Lo\'eve (KL) expansion} \cite{ghanem2003stochastic}. The idea behind the truncated KL expansion is that a spectral decomposition of the uncertain parameters can be used to express them as a linear combination of an infinite number of iid (independent and identically distributed) random variables, following which the infinite series can be truncated by picking basis functions corresponding to the highest ranked eigenvalues by magnitude. 
This is the functional analogue of the well-known principal component analysis (PCA) \cite{jolliffe2011principal} used extensively in the machine learning (ML) community.
Although extensively used, the truncated KL expansion (or PCA) only considers information contained in the input data resulting in an overestimation of the intrinsic dimensionality of the stochastic parameter space. 

Generally speaking, an effective technique for dimensionality reduction needs to exploit intrinsic structure within the underlying map being approximated. 
One such technique that has been recently popularized is the method of \textit{active subspaces} (AS), introduced in \cite{constantine2014active}. 
An AS is a low-dimensional linear manifold embedded within the true high-dimensional parameter space, which maximally captures the variation of the underlying map. 
AS has been successfully applied to numerous applications since it's introduction \cite{gilbert2016global, constantine2015exploiting, leon2017identifiability, grey2018active}. 
However, a key drawback of the original AS framework is it's reliance on gradient information about the model outputs, which are often difficult (if not impossible) to obtain. 
To overcome the gradient requirement for AS recovery, \cite{tripathy2016gaussian} proposed a framework which subsumes the AS projection matrix into the covariance kernel in GP regression and attempt to learn it from available data. 

In this work, we propose a simple solution for recovering the AS and constructing surrogate models that does not rely on non-Euclidean algorithms for surrogate model optimization. 
Specifically, we express the AS projection matrix as the Gram-Schmidt orthogonalization of an unconstrained matrix. 
We rely on the fact that the Gram-Schmidt procedure is fully differentiable (and thus perfectly amenable to gradient-based learning through backpropagation). 
Furthermore, our method is agnostic to the specific class of function approximation. This is contrasted with the previous gradient-free AS framework proposed in \cite{tripathy2016gaussian}. 
Lastly, we couple this idea with deep neural networks (DNNs) and demonstrate true AS recovery.
An added benefit of the proposed approach is that one can use it to improve DNN identifiability (even though it is not our primary concern). 

This manuscript is organized as follows. 
We begin with a formal description of the UP problem and the surrogate approach to solving it in Sec. \ref{sec:up}. 
We then, in Sec. \ref{sec:methodology}, review the classical approach to AS recovery (see Sec. \ref{sec:as_classic}) and the gradient-free GPR based AS approach proposed in \cite{tripathy2016gaussian} (see Sec. \ref{sec:as_grad_free}). 
This is followed up with the description of our approach to AS, in Sec. \ref{sec:deepas}. 
Finally, we demonstrate the proposed methodology on challenging high-dimensional surrogate modeling problems in Sec. \ref{sec:examples} and demonstrate that we recover the true AS through comparisons with the classical approach.

\section{METHODOLOGY}
\label{sec:methodology}

\subsection{The surrogate approach to uncertainty propagation}
\label{sec:up}
Consider a physical system modeled with a (potentially complex, coupled) system of partial differential equations.
The PDE(s) is solved numerically using a black-box computer code, which we denote as $f$. 
$f$ may be thought of as a multivariate function which accepts a vector of inputs $\bxi \in \bXi \subset \R^D$ and produces a scalar quantity of interest (QoI) $f(\bxi) \in \mathcal{Y} \subset \R$. 
Information about $f$ may be obtained through querying the solver at suitable input design locations $\bxi$. 
We allow for the possibility that our measurement from the computer code may be noisy, i.e., $y = f(\bxi) + \epsilon$, where $\epsilon$ is a random variable (the measurement noise might arise as a consequence of quasi-random stochasticity or chaotic behavior). 
Given this setup, the uncertainty propagation (UP) task is summarized as follows. Given a formal description of the uncertainty in the input parameters, $\bxi \sim p(\bxi)$, we would like to estimate the statistical properties of the QoI. This includes, the probability density, 
\begin{equation}
    \label{eqn:qoi_pdf}
    p(f) = \int \delta(f - f(\bxi)) p(\bxi) \mathrm{d}\bxi, 
\end{equation}
and measures of central tendency such as the mean:
\begin{equation}
    \label{eqn:qoi_mean}
    \mu_f = \int f(\bxi) p(\bxi) \mathrm{d}\bxi,
\end{equation}

and variance:
\begin{equation}
    \label{eqn:qoi_var}
    \sigma_{f}^{2} = \int (f(\bxi) - \mu_f )^2 p(\bxi) \mathrm{d}\bxi,
\end{equation}
where, $\delta(\cdot)$ in Eqn. (\ref{eqn:qoi_pdf}) refers to the Dirac $\delta$-function. 

As already discussed in the introduction, the standard MC method is infeasible when there is a large computational cost associated with querying $f$ and one must resort to the surrogate approach - replacing the true simulator $f$ with an accurate, cheap-to-evaluate approximation, $\hat{f}$. 
To do this, one queries $f$ at a set of $N$ carefully selected design locations $\X = (\bxi^{i})_{i=1}^{N}$, resulting in a corresponding set of measurements, $\y = (y^{i})_{i=1}^{N}$. We refer to the observed data, collectively, as $\calD = \{\X, \y \}$. Although the task of careful selection of the input design locations are a subject of a great deal of research, an in-depth discussion of this topic is beyond the scope of the present work. Here we simply assume that we are given $\calD$. 

\subsection{Active subspaces}
\label{sec:as}
The fact that we are working in a high-dimensional regime $D (\gg 1)$ makes the task of constructing an accurate surrogate model with limited data $\left(N \approx \mathcal{O}(D) \right)$ practically infeasible because of the curse of dimensionality. To circumvent this, one seeks to exploit low-rank structure within the true response $f$ and methods for doing so are broadly categorized as `dimensionality reduction' techniques.  
In this work, we focus on the case where the response admits the following structure: 
\begin{equation}
    \label{eqn:asstructure}
    f(\bxi) = g(\bzeta) = g(\W^T \bxi), 
\end{equation}
where, $\W \in \R^{D \times d}$ is a tall-and-skinny matrix of orthogonal columns which projects the high-dimensional input $\bxi$ to $\bzeta$ lying in a $d$-dimensional subspace such that $d \ll D$. 
In particular, $\W$ is constrained to be an element of the set:
\begin{equation}
    \label{eqn:stiefel}
    \mathcal{V}_{d}\left(\R^D \right) = \left\{ \mathbf{A} \in \R^{D \times d} : \mathbf{A}^T \mathbf{A} = \mathbf{I}_d \right\}.
\end{equation}
$\mathcal{V}_{d}\left(\R^D \right)$ is known as the \textit{Stiefel manifold} with $\mathbf{I}_d$ being the identity matrix in $\R^{d \times d}$ and $g:\R^d \rightarrow \R$ is known as the \textit{link function}.
The structure posited in Eqn. (\ref{eqn:asstructure}) takes on physical meaning where the columns of $\W$ correspond to directions of the input space most sensitive to variation in the response $f$. 
The dimensionality reduction induced by the introduction of this structure, significantly simplifies the task of learning an accurate surrogate model. 

\subsubsection{Classical approach to active subspaces}
\label{sec:as_classic}
The classical approach to recovering the \textit{active subspace}, introduced in \cite{constantine2014active}, proceeds as follows. 
Let the gradient of the QoI w.r.t. the input be denoted as $\nabla_{\bxi} f = \left(\frac{\partial f}{\partial \bxi_1}, \frac{\partial f}{\partial \bxi_2}, \cdots, \frac{\partial f}{\partial \bxi_D} \right) \in \R^D$.
Given a probability distribution $\rho$ endowed upon the input space, we define the symmetric positive semi-definite matrix, 
\begin{equation}
    \label{eqn:as_matrix_true}
    \mathbf{C} = \int (\nabla_{\bxi}f(\bxi)) (\nabla_{\bxi}f(\bxi))^T \rho(\bxi) \mathrm{d}\bxi,
\end{equation}
which admits the spectral decomposition $\mathbf{C} = \mathbf{V} \boldsymbol{\Lambda} \mathbf{V}^T$, where $\boldsymbol{\Lambda}$ is a diagonal matrix of eigenvalues ordered by magnitude. Separating the $d$ largest eigenvalues from the rest, we can write the matrix of eigenvectors, $\mathbf{V}$, as:
\begin{equation}
    \mathbf{V} = \left[\mathbf{V}_1, \mathbf{V}_2 \right],
\end{equation}
where, $\mathbf{V}_1 \in \R^{D \times d}$ is a matrix consisting of the eigenvectors corresponding to the $d$ largest eigenvalues and $\mathbf{V}_2 \in \R^{D \times (D-d)}$ is composed of the remaining eigenvectors. The active subspace projection matrix, then, is simply, $\W = \mathbf{V}_1$. 
Since the integral in Eqn. (\ref{eqn:as_matrix_true}) is intractable analytically (due to the black-box nature of $f$), one only has access to discrete samples of the gradient at input locations $\bxi$ sampled from the distribution $\rho$. Given a dataset of $S$ gradient evaluations, $\mathbf{g}^{(i)} = \nabla_{\bxi} f(\bxi^{(i)}), i = 1, 2, \cdots, S$, where the $\bxi^{(i)}$s are sampled iid from $\rho$, an approximation to the matrix $\mathbf{C}$ may be constructed as:
\begin{equation}
    \label{eqn:as_matrix_approx}
    \mathbf{C}_S = \frac{1}{S} \sum_{i=1}^{S} \mathbf{g}^{(i)} \mathbf{g}^{(i),T}.
\end{equation}
One may think of the approximation $\mathbf{C}_S$ as an empirical covariance matrix. After recovering the projection matrix $\W$ through the above procedure, one can obtain projected inputs, using $\z = \W^T \bx$ and using a suitable technique such as Kriging to learn the link function $g(\cdot)$. 

\subsubsection{Gradient-free approach to active subspaces}
\label{sec:as_grad_free}
As discussed in Sec. \ref{sec:as_classic}, the classic approach to AS recovery requires the evaluation of an empirical covariance matrix from samples of the gradient $\nabla_{\bxi}f$. Obtaining gradient samples in challenging in practice. 
In some cases (such as simple dynamical systems), one might have access to an adjoint solver which can compute the gradients of the QoI wrt the input parameters \cite{jameson2003aerodynamic}. 
In other cases, the gradients can be approximated through finite differences (FD). Note that a single first-order FD gradient evaluation requires 2 expensive forward model runs. 
Lastly, one might even approximate gradients through approximate global models for the data \cite{jefferson2015active}. 
In general, the black-box nature of the response as well the associated cost of FD gradients means that one simply does not have access to $\nabla_{\bxi}f$ and therefore cannot construct $\W$ through the classical approach. 
To alleviate this limitation, \cite{tripathy2016gaussian} introduced a methodology for constructing surrogate models without requiring gradient information. The gradient-free approach relies on two key ideas:
\begin{enumerate}
\item In GPR, prior knowledge about the underlying function can be encoded in a principled manner through the mean and the covariance functions of the GP. Thus, a new covariance kernel may be defined where the AS projection matrix $\W$ is simply a hyperparameter and learned through available data, $\calD$. Formally, the prior knowledge about the active subspace structure described in Eqn. (\ref{eqn:asstructure}) is expressed through a GP kernel which takes on the form:
\begin{equation}
    \label{eqn:as_kernel}
    k_{\mathrm{AS}}(\bx, \bx') = k_{\mathrm{base}}(\z, \z') = k_{\mathrm{base}}(\W^T \bx, \W^T \bx'), 
\end{equation}
where, $k_{\mathrm{base}}(\cdot, \cdot)$ is any standard kernel (such as the \textit{Matern} or \textit{Radial basis function (RBF)} kernels) which expresses prior knowledge about the regularity properties of the link function $g(\cdot)$. 
Once the active subspace kernel has been defined, inference in GPR proceeds through the maximization of the log marginal likelihood of the data wrt the kernel hyperparameters i.e., 
\begin{equation}
    \label{eqn:as_grad_free_opt}
    \W^*, \mathcal{H}^*, \sigma^{*}_{n} = \underset{\W, \mathcal{H}}{\mathrm{argmax}} \log p(\y|\X, \W, \mathcal{H}, \sigma_{n}),
\end{equation}
where, $\mathcal{H}$ is the set of all hyperparameters of the base kernel $k_{\mathrm{base}}$, and $\sigma_{n}$ is the standard deviation of the likelihood noise. 

\item While it is easy to enforce positivity constraints on the hyperparameters $(\mathcal{H}, \sigma_{n}) = \bphi$, the optimization task in Eqn. (\ref{eqn:as_grad_free_opt}) is made challenging because of the fact that it is non-trivial to enforce the orthogonality constraints on the projection matrix $\W$. 
In order to do so, the complete methodology of \cite{tripathy2016gaussian} relies on a coordinate-ascent scheme to iteratively optimize over the variables $\bphi$ while keeping $\W$ constant and vice versa. 
The optimization steps over $\bphi$ proceed via standard second-order techniques for unconstrained optimization, such as the L-BFGS method \cite{byrd1995limited}. 
The optimization steps over the projection matrix $\W$ utilize an adapted version of gradient-ascent on the Stiefel manifold described in \cite{wen2013feasible}.  
\end{enumerate}

\subsection{Deep active subspaces}
\label{sec:deepas}
The methodology introduced by \cite{tripathy2016gaussian} lifts the gradient requirement of the classical approach to AS recovery by subsuming the AS projection matrix into the covariance kernel of a GP. While the methodology is sound and experimentally shown to recover the true AS, it suffers from two major drawbacks - 
\begin{enumerate}
    \item It is not agnostic to the choice of the surrogate model. 
    Note that the gradient-free method described in Sec. \ref{sec:as_grad_free} necessitates a GP surrogate by construction. 
    Inspite of the elegance of GPR, arising out of the principled framework it offers for incorporating prior knowledge, quantifying epistemic uncertainty and performing model selection, it's standard formulation scales poorly due to the $\mathcal{O}(N^3)$ inversion of the (potentially dense) covariance matrix required at each optimization step. 
    While sparse GPR \cite{titsias2009variational, snelson2006sparse} partially alleviates this poor scaling through the introduction of $M (\ll N)$ inducing variables or `pseudo-inputs', the task of selecting or optimizing for the inducing input locations is non-trivial. 
    \item The proposed solution for optimizing over the projection matrix $\W$, while respecting orthogonality constraints, is itself non-trivial, introduces $Dd$ additional hyperparameters into the covariance kernel, and is prone to getting trapped in local stationary points \cite{tripathy2016gaussian}. 
\end{enumerate}
We propose, here, a much simpler approach that is:
\begin{enumerate}
    \item Is agnostic to the choice of the link function approximator, 
    \item Is trivial to implement. 
\end{enumerate}

Specifically, we express $\W$ as:
\begin{equation}
    \label{eqn:W_express}
    \W = h(\mathbf{Q}),
\end{equation}
where, $\mathbf{Q} \in \R^{D \times d}$ lies on the standard Euclidean space, and $h:\R^{D \times d} \rightarrow \R^{D \times d}$ orthogonalizes the columns of $\mathbf{Q}$. Specifically, we chose $h$ to be the celebrated \textit{Gram-Schmidt (GS) orthonormalization} procedure \cite{bjorck1967solving}. The GS process may be summarized as follows. Given an unconstrained matrix $\mathbf{Q} = [\mathbf{q}_1, \mathbf{q}_2, \cdots, \mathbf{q}_d] \in \R^{D \times d}$, where the $\mathbf{q}_i$s are the columns of $\mathbf{Q}$, we apply the transformation, 
\begin{equation}
    \label{eqn:gs_trans}
    \mathbf{w}_i = \mathbf{w}_{i-1} - \left( \frac{\mathbf{w}_{i-1}^{T} \mathbf{q}_{i}}{\mathbf{w}_{i-1}^{T} \mathbf{w}_{i-1}} \right) \mathbf{w}_{i-1},\ i=2, 3, \cdots, d,
\end{equation}
with $\mathbf{w}_1 = \mathbf{q}_1$. The projection matrix $\W$ is then assembled by normalizing the $\mathbf{w}_i$s, i.e., $\W = \left[\frac{\mathbf{w}_1}{\| \mathbf{w}_1 \|_2}, \frac{\mathbf{w}_2}{\| \mathbf{w}_2 \|_2}, \cdots, \frac{\mathbf{w}_d}{\| \mathbf{w}_d \|_2}\right]$.

Now one only needs to care about the the Euclidean matrix $\mathbf{Q}$, and optimize it to the available data. 
Noting that the transformation specified by Eqn. (\ref{eqn:gs_trans}) is fully differentiable (as it composed entirely of differentiable mathematical operations), one may simply define a routine implementing the GS process using a backpropagation-capable library (such as \texttt{TensorFlow} or \texttt{PyTorch}) to obtain exact gradients of any QoI wrt $\mathbf{Q}$. 

Since the projection matrix $\W$ has been reparameterized without any concern for the specific structure of the link function, $g$, we are free to pick any suitable class of function approximator for $g$. In this work, we define $g$ to be a deep neural networks (DNN) \cite{goodfellow2016deep}, a class of highly flexible nonlinear function approximators with satisfy universal approximation properties. Formally, a $L$-layered DNN representation for $g$ is defined as:
\begin{equation}
    \label{eqn:link_dnn}
    g(\z) = f_{L+1} \circ f_{L} \circ \cdots \circ f_1(\z),
\end{equation}
where, $f_i(\z_{i-1}) = h_i(\W_i^T \z_{i-1} + \mathbf{b}_i)$, with $\W_i \in \R^{d_{i} \times d_{i-1}}, \mathbf{b}_i \in \R^{d_{i}}, \z_i = \R^{d_{i}}, \z_0 = \z, \z_{L+1} = g(\z)$, and $h_i(\cdot)$ is a suitable nonlinear function applied elementwise on it's argument. 
$h_{L+1}$ is set to be the identity function (since we are dealing with unconstrained real-valued outputs) and the  other $h_i$s are set as the hyperbolic tangent function, a standard choice in the literature.
The matrices $\W_i$s and the vectors $\mathbf{b}_i$s are called the `weights' and `biases' of the DNN and here we denote all of them collectively as $\btheta = \{\W_1, \W_2, \cdots, \W_{L+1}, \mathbf{b}_1, \mathbf{b}_2, \cdots, \mathbf{b}_{L+1} \}$. 
The full surrogate, is there expressed as:
\begin{equation}
    \label{eqn:full_surr}
    \hat{f}(\bxi; \btheta) = g(h(\mathbf{Q})^T \mathbf{\bxi}; \btheta),
\end{equation}
where the unknown parameters $(\btheta, \mathbf{Q})$ can be optimized through standard gradient-descent techniques. In this work, we use the famous Adaptive Moments (ADAM) optimization method \cite{kingma2014adam}.

\section{NUMERICAL EXAMPLES}
\label{sec:examples}

\subsection{Synthetic example with known active subspace}
\label{sec:synthetic}
Let $f:\R^D \rightarrow \R$ such that $f(\bxi) = g(\bzeta) = g(\W^T \bxi)$ where $\W \in \mathcal{V}_{d}\left(\R^D \right)$. Define $g: \R^d \rightarrow \R$ as a quadratic function in $\R^d$:
\begin{equation}
\label{eqn:link_1d_as}
g(\bzeta) = \W^T \bxi = \alpha + \bbeta^T \z + \z^T \bGamma \z.
\end{equation}
The gradients of $f$ are given by
\begin{equation}
    \label{eqn:f_grad}
    \nabla f(\bxi) = \left(\beta + 2\bxi^T \W \bGamma \right) \W^T.
\end{equation}

For this pedagogical example, we set $D = 20$ and test our approach on two cases with true AS dimensionality, $d = 1$ and $2$. The data for inputs $\bxi$, $\alpha$, $\beta$ and $\bGamma$ are generated by sampling standard Gaussians of appropriate shapes. The matrix $\W$ is generated by performing the QR decomposition on a similarly generated matrix in $\R^{D \times d}$. Random seed is fixed for reproducibility. The output data $\y$ used for training is standardized, i.e. it is scaled to have 0 mean and unit variance. 

\subsubsection{Case 1: 1 dimensional active subspace}
\label{sec:as_1d}
\begin{figure}[ht]
    \centering
    \includegraphics[width=0.5\textwidth]{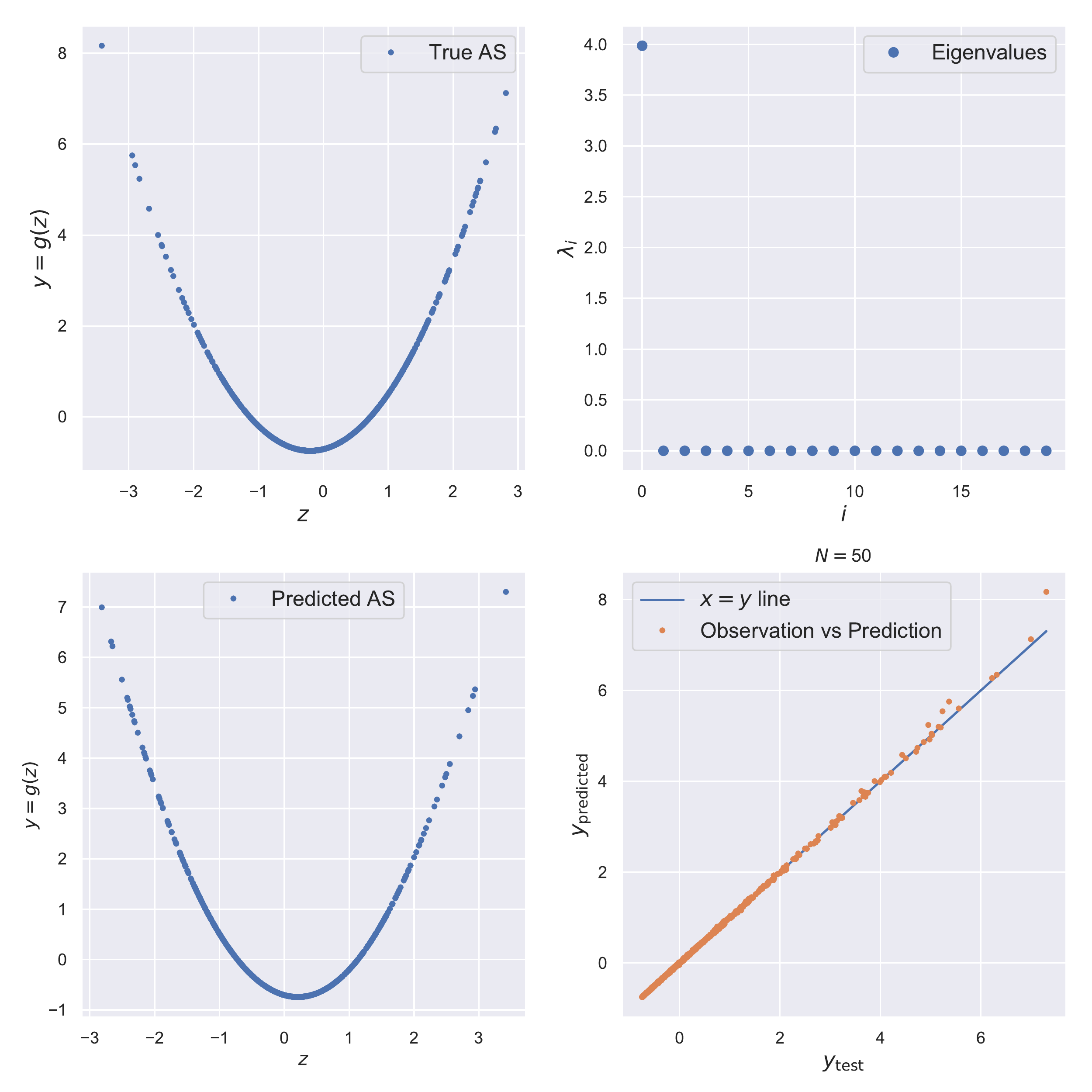}
    \caption{Synthetic function with $D=20$ input dimensions admitting an $d=1$ dimensional active subspace. Top left - True AS of $f$. Bottom left - AS predicted by DNN. Top right - Spectral decomposition of the empirical covariance of the gradients. Bottom right - Comparison of predicted output and correct output on the test dataset.}
    \label{fig:ex1_case_1_N_50}
\end{figure}

We begin testing our proposed gradient-free approach on a synthetic function which an AS of dimensionality $d=1$.
To train the AS DNN, we use $N=50$ input-output observations. 
Furthermore, the output data is corrupted with zero mean Gaussian noise of standard deviation $1\times 10^{-2}$.
The link function is approximated with a 2 layer DNN of 50 units per hidden layer and $L_2$ regularization with a weight decay constant of $1 \times 10^{-4}$ is used to prevent overfitting. 
The ADAM optimizer is set to perform $3 \times 10^4$ iterations with a base learning rate of $1\times 10{-3}$ dropped by a factor of $10$ every $10^{4}$ iterations.
In Fig. \ref{fig:ex1_case_1_N_50} we visually compare the true AS for this case and the AS discovered by the DNN. 
We note that they are very close, indicating that our approach has found the correct AS. 
Fig. \ref{fig:ex1_case_1_N_50} also shows a comparison of the AS DNN predicted response with the true response from a test dataset of $500$ observations. 
Qualitatively, the predictions match the observations very closely. Quantitatively, we achieve a root mean square error of $0.039689$ on the test dataset. 
Note that we pursue no further optimization of our DNN structure. 

\subsection{Case 2: 2 dimensional active subspace}
\label{sec:as_2d}
\begin{figure}[ht]
    \centering
    \includegraphics[width=0.5\textwidth]{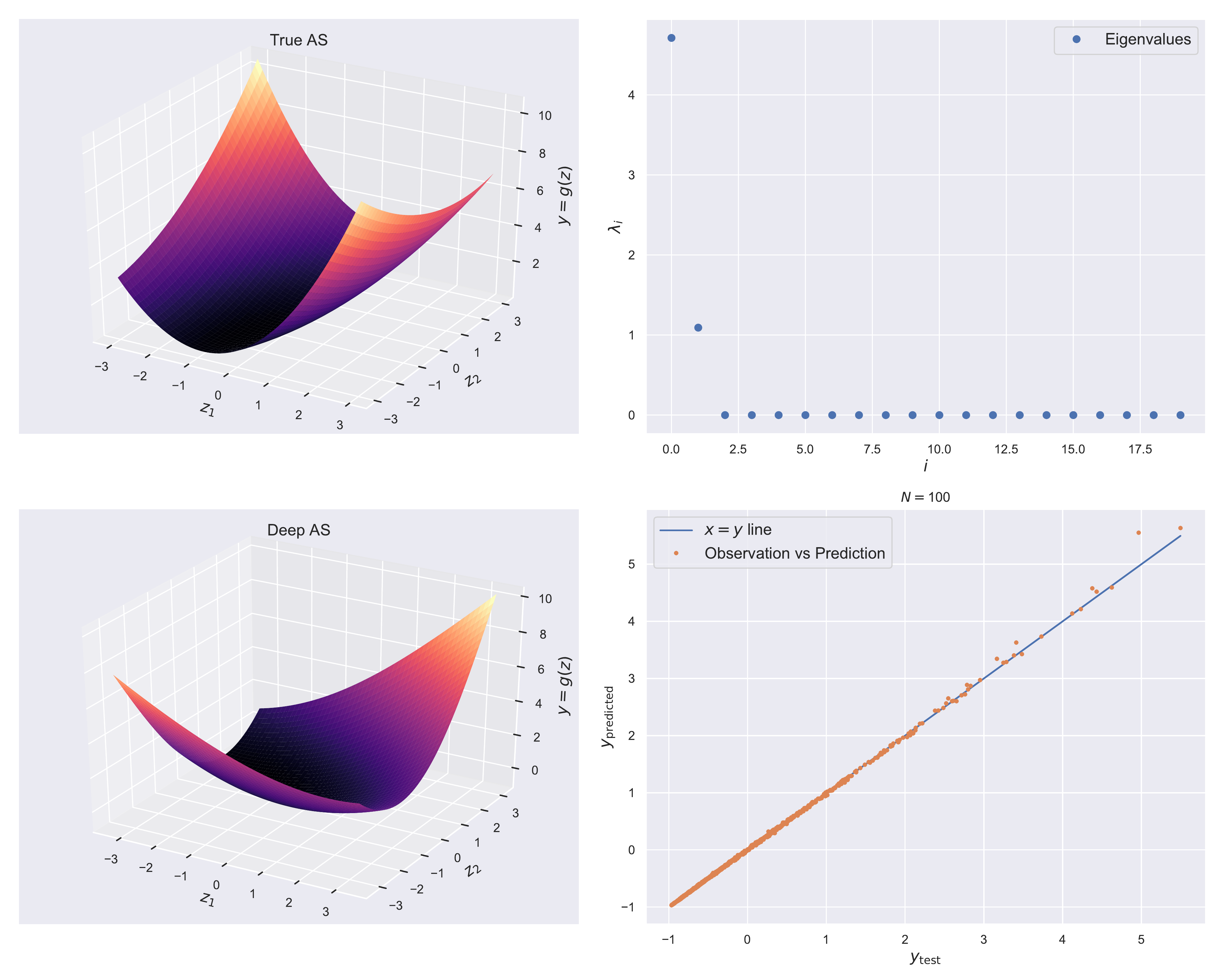}
    \caption{Synthetic function with $D=20$ input dimensions admitting an $d=2$ dimensional active subspace. Top left - True AS of $f$. Bottom left - AS predicted by DNN. Top right - Spectral decomposition of the empirical covariance of the gradients. Bottom right - Comparison of predicted output and correct output on the test dataset.}
    \label{fig:ex1_case_2_N_100}
\end{figure}

We now test our proposed on a synthetic function with AS dimensionality, $d=2$.
We use $N=100$ input-output observations for training and corrupt the output data with zero mean Gaussian noise of standard deviation $1\times 10^{-2}$.
We retain all other experimental settings from Sec. \ref{sec:as_1d}.
A comparison of the true AS and the predicted AS shown in Fig. \ref{fig:ex1_case_2_N_100} reveals that we recover the low-dimensional quadratic response upto arbitrary rotations of the coordinate system. 
We compare the predicted response of the AS DNN and true outputs from a test dataset of $500$ observations. Again, we obtain excellent qualitative agreement as seen in Fig. \ref{fig:ex1_case_2_N_100} and quantitatively, we obtain a root mean squared error of $0.028748$ between the predicted and true outputs. 

\subsection{Benchmark elliptic PDE example}
\label{sec:constantine_spde}
Consider the following stochastic elliptic partial differential equation defined on the unit square in $\R^2$:
\begin{equation}
    \nabla \cdot \left(a(\mathbf{s}) \nabla u(\mathbf{s}) \right) = 1, \mathbf{s} \in \Omega = [0, 1]^2,
\end{equation}
with boundary conditions:
\begin{align}
    \label{eqn:const_bc}
    u(\mathbf{s}) &= 0, \mathbf{s} \in \Gamma_u, \\
    \nabla u(\mathbf{s}) \cdot \mathbf{n} &= 0, \mathbf{s} \in \Gamma_n, 
\end{align}
where, $\Gamma_u$ is the top, bottom and left boundaries and $\Gamma_n$ denotes the right boundary of $\Omega$. 
The diffusion coefficient $a$ (or conductivity field) is a spatially-varying uncertain input, and it's logarithm is modeled as a 0-mean Gaussian random field, i.e., $\log a(\mathbf{s}) \sim GP(a|0, k(\mathbf{s}, \mathbf{s}')))$, where, the covariance function $k$ is defined as follows:
\begin{equation}
    \label{eqn:const_elliptic_cov}
    k(\mathbf{s}, \mathbf{s}') = \exp \left(- \frac{\sum_{i=1}^{2} |\mathbf{s}_i - \mathbf{s}_{i}^{'}|}{\ell} \right),
\end{equation}
with $\ell$ being the correlation length. This formalization of the uncertainty around $a(\mathbf{s})$ makes it a stochastic process - an infinite dimensional quantity. 
We use the truncated KL expansion to perform a preliminary dimensionality reduction by expressing the logarithm of the field as:
\begin{equation}
    \label{eqn:ex_2_log_a}
    \log a(\mathbf{s}) = \sum_{i=1}^{100} \sqrt{\lambda_i} \varphi_i (\mathbf{s}) x_i, 
\end{equation}
where, the $\lambda_i$s and the $\varphi_i$s are the eigenvalues and eigenfunctions of the correlation function, numerically obtained using the \textit{Nystr\"om approximation} \cite{bilionis2016bayesian}, and the $x_i$s are uncorrelated, standard normal random variables.
Denote all the $x_i$s collectively as $\mathbf{x} = (x_1, x_2, \cdots, x_{100}) \sim \calN(\bx|\mathbf{0}, \mathbf{I}_{100})$. 
We are interested in the following scalar QoI:
\begin{equation}
    \label{eqn:ex2_qoi}
    q(\mathbf{x}) = \mathcal{F}[u(\mathbf{s}; \mathbf{x})] = \frac{1}{|\Gamma_2|}\int u (\mathbf{s}; \mathbf{x})  \mathrm{d}\mathbf{s}.
\end{equation}
Given a realization of the random variable, $\mathbf{x} = (x_1, x_2, \cdots, x_100) $, one can generate a realization of the QoI, $q$, whose statistics one wishes to estimate. 
We have, at our disposal, a dataset of $300$ realizations of the random variable $\bx$ and the corresponding solution $q$. 
With this dataset, we construct a surrogate that maps $\bx$ to $q$, i.e., $\hat{f}:\R^{300} \rightarrow \R$. 
We will consider two cases of the correlation length $\ell$ - a short correlation length of $\ell = 0.01$ and a long correlation length of $\ell = 1$ and attempt to recover as AS with $d=1$. 
We randomly shuffle and split the data into a training set of $250$ samples, and test on the remaining $50$ samples. 
The output data is standardized to have zero mean and unit variance for numerical stability.
The dataset for this example, and the code to generate it, can be found here: \href{https://github.com/paulcon/as-data-sets/tree/master/Elliptic_PDE}{https://github.com/paulcon/as-data-sets/tree/master/Elliptic\_PDE}.
Once again, we set our approximation of the link function to be a DNN with 2 hidden layers and 50 units per layer. 
All other experimental settings from Sec. \ref{sec:synthetic} are retained. 
Lastly, for this example, samples of the QoI gradients are available and we use this to compare our results with the results obtained from classic AS. 
For the case of the classic AS, we use GPR as the link function.

\begin{figure}[ht]
    \centering
    \includegraphics[width=0.5\textwidth]{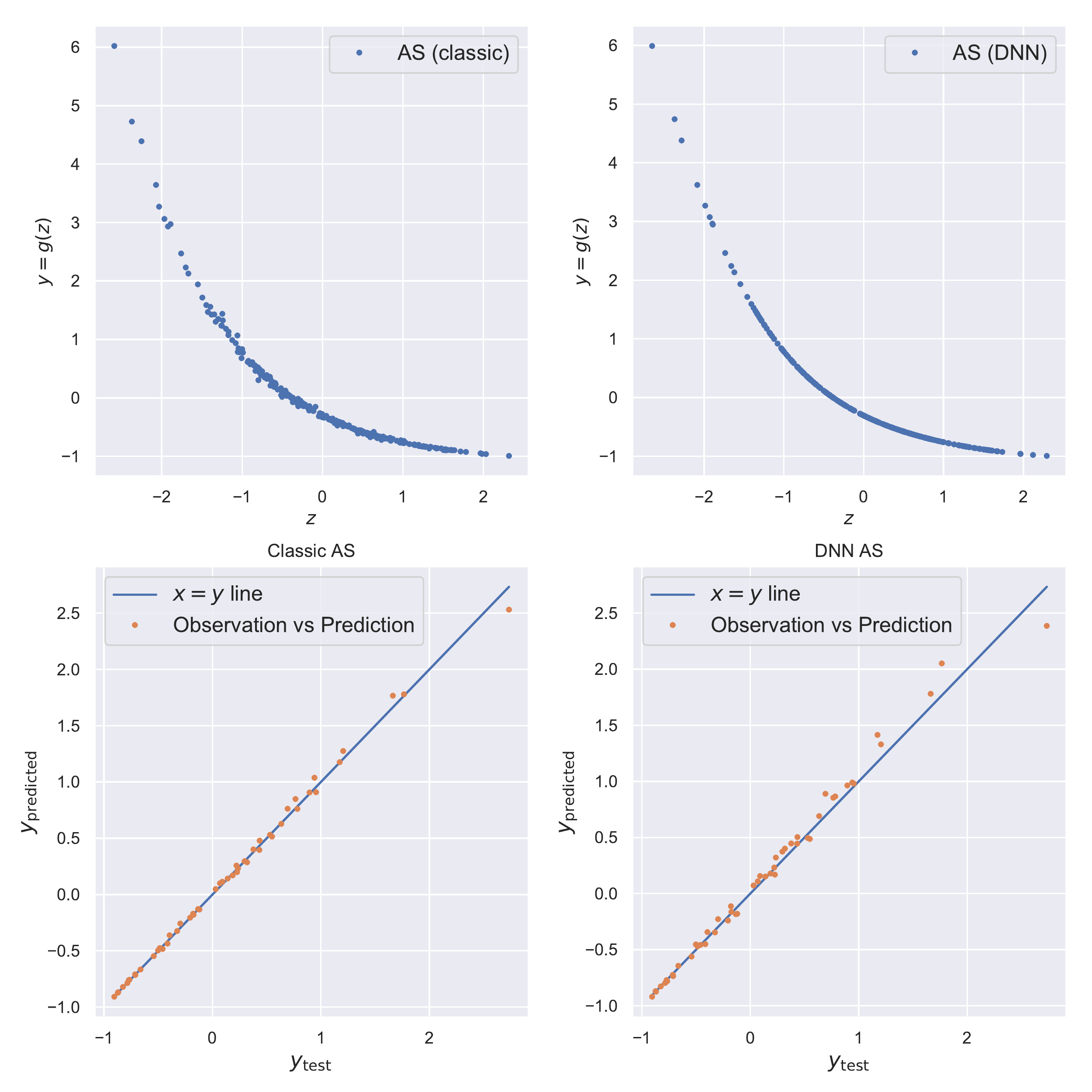}
    \caption{Stochastic elliptic PDE with $\ell = 1$ - The plots on the top visualize the 1d active subspace recovered by our gradient-free DNN AS approach and the classic AS approach. 
    The bottom plots compare the output predictions vs observations on the test dataset for the DNN AS and the classic AS approaches.}
    \label{fig:ex_2_long_corr}
\end{figure}

\begin{figure}[ht]
    \centering
    \includegraphics[width=0.5\textwidth]{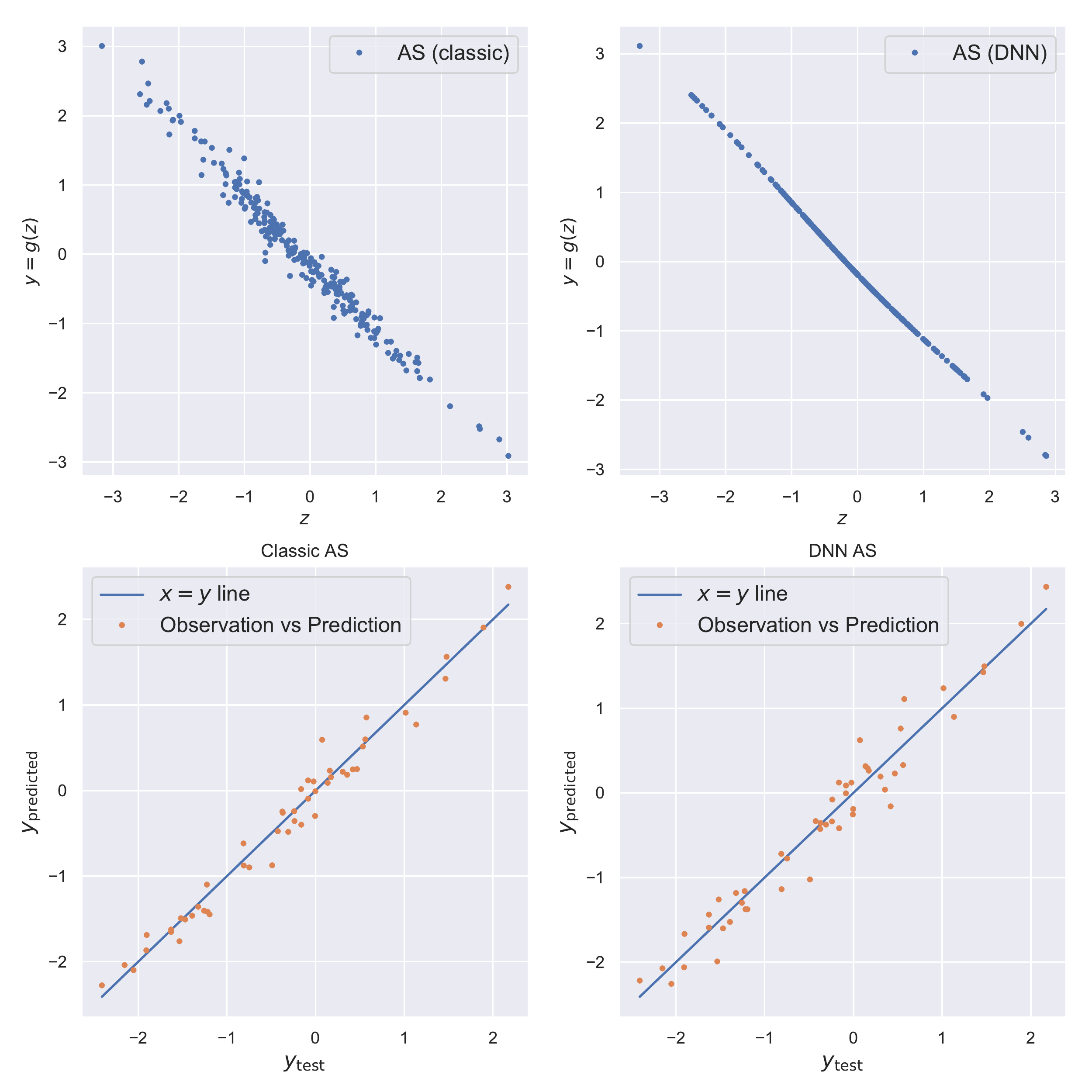}
    \caption{Stochastic elliptic PDE with $\ell = 0.01$ - The plots on the top visualize the 1d active subspace recovered by our gradient-free DNN AS approach and the classic AS approach. 
    The bottom plots compare the output predictions vs observations on the test dataset for the DNN AS and the classic AS approaches.}
    \label{fig:ex_2_short_corr}
\end{figure}

Fig. \ref{fig:ex_2_long_corr} shows results comparing the deep AS approach and the classic AS approach for the $\ell = 1$ case. 
Fig. \ref{fig:ex_2_short_corr} shows the same for the case of $\ell = 0.01$. 
We observe that there is good qualitative agreement between the AS recovered by gradient-free deep AS approach and the gradient-based classic approach. 
This serves to verify the fact that our approach does indeed recover the correct AS. 
Table \ref{tab:elliptic_pde_rmse} shows a comparison of the RMSE error in the prediction of the outputs from the test dataset, for both $\ell$ cases. 
We observe that inspite of the fact that we do not use the information about the gradients of QoI, our gradient-free deep AS approach is are able to achieve RMSE comparable to the classic AS.
Once again, we emphasize that we do not pursue optimization of the modeling choices involved in the DNN approximation of the link function. 
\begin{table}[ht]
    \centering
    \begin{tabular}{||c|c|c||} 
        \hline \hline
          Approach & $\ell = 1$ & $\ell = 0.01$ \\ 
        \hline \hline
         Classic AS & 0.04378 & 0.17276 \\
        \hline
         Deep AS & 0.09241 & 0.23626 \\
        \hline \hline
    \end{tabular}
    \caption{Root mean square error (RMSE) on test dataset predictions from classic AS and deep AS response surfaces.}
    \label{tab:elliptic_pde_rmse}
\end{table}

An interesting observation that emerges from the comparison of the classic and deep AS approaches for the short correlation length in Fig. \ref{fig:ex_2_short_corr} is that inspite of recovering a one-dimensional AS, the test data predictions from both approaches show a discrepancy from their true values. 
Since the QoI is generated from a deterministic computer code, we cannot explain this deviation as `noise'. 
Rather, this suggests that a linear dimensionality reduction is sub-optimal and one might wish to recover a nonlinear generalization of the active subspace, such as the one discussed in \cite{tripathy2018deep}. 
An investigation into this shortcoming is beyond the scope of the present work.
Finally, one may note that both the classic and the deep AS approaches perform much better on the $\ell = 1$ case, relative to the $\ell = 0.01$ case. 
This is unsurprising, considering that it becomes much more difficult to capture the uncertainty of the diffusion coefficient as it's lengthscale reduces.

\section{Conclusion}
\label{sec:conclusion}
In this work, we presented a novel methodology for recovering active subspaces (AS) and constructing surrogate models in applications with high-dimensional uncertain parameter spaces. 
Our approach rests on a reparameterization of the AS projection matrix using the Gram-Schmidt procedure. 
Noting the fact that the GS procedure is, in principle, a fully-differentiable operation, one can easily backpropagate through the GS process to obtain gradients necessary in standard optimization routines. 
This formulation liberates us from the GPR approach of past gradient-free AS methods, and allows us to couple AS recovery with deep neural networks (DNNs) - a nonlinear function approximator which can be scaled to high-dimensions/larger datasets much more easily. 
We demonstrated the proposed approach on benchmark problems in AS recovery, showing that we do indeed recover the correct AS. 

This work represents a first-step toward scaling gradient-free recovery of AS - an important objective since many quantities of interest encoding physical laws exhibit AS or AS-like ridge structure \cite{constantine2016many}. 
Our long term interest is in the development of efficient, Bayesian surrogates that are capable of quantify epistemic uncertainty. 
The reparameterization of the projection matrix allows us to leverage standard Bayesian inference methods such as stochastic variational inference (SVI) \cite{hoffman2013stochastic} or Hamiltonian Monte Carlo (HMC) \cite{hoffman2014no} to construct Bayesian AS surrogates without resorting to specialized Riemannian manifold versions of these techniques. 







\bibliographystyle{asmems4.bst}


%
% ...

\bibliography{bibliography.bib}



\end{document}